\documentclass[final,3p,twocolumn,times,sort&compress]{elsarticle}
\usepackage{float, graphics, graphicx, epsfig, bm, amsmath, amsthm, amssymb, amsfonts,multirow}
\usepackage[position=top, subrefformat=parens]{subfig}
\usepackage[unicode, colorlinks=true, urlcolor=blue, linkcolor=blue, citecolor=blue, pdfauthor={ }, pdftitle={ }, pdfsubject={ }, pdfkeywords={ }]{hyperref}
\usepackage{titlesec}
\titleformat{\subsubsection}[block]{\centering\itshape}{\thesubsubsection.}{1em}{}
\renewcommand{\thesubsubsection}{\alph{subsubsection}}

\usepackage{pdfpages}

\journal{Optical Materials}

\begin{document}
\begin{frontmatter}

\title{Lanthanide-Dependent Clustering in Yb$^{3+}$/Ln$^{3+}$ Co-Doped CaF$_2$ Nanocrystals: Correlating Spectroscopic Signatures with DFT Insights}

\author[UC,DW,FSU]{Sangeetha Balabhadra\corref{co}}

\author[USTC]{Haoming Xu\corref{co}}

\author[HUT,USTC]{Jiajia Cai\corref{cor}}
\ead{jjcai@mail.ustc.edu.cn}

\author[USTC]{Chang-Kui Duan}

\author[UC,DW]{Michael F.\ Reid\corref{cor}}
\ead{mike.reid@canterbury.ac.nz}

\author[UC,DW]{Jon-Paul R.\ Wells\corref{cor}}
\ead{jon-paul.wells@canterbury.ac.nz}

\affiliation[UC]{
	organization={School of Physical and Chemical Sciences}, 
	addressline={University of Canterbury}, 
	city={Christchurch}, 
	postcode={8140}, 
	state={PB 4800}, 
	country={New Zealand}
}

\affiliation[DW]{
	organization={The Dodd-Walls Centre for Photonic and Quantum Technologies}, 
	country={New Zealand}
}

\affiliation[USTC]{
	organization={Laboratory of Spin Magnetic Resonance, and School of Physical Sciences}, 
	addressline={University of Science and Technology of China}, 
	city={Hefei},  
	postcode={230026}, 
	state={Anhui}, 
	country={China}
}

\affiliation[FSU]{
	organization={Department of Chemistry, Physics \& Materials Science}, 
	addressline={Fayetteville State University}, 
	city={Fayetteville}, 
	postcode={28301}, 
	state={NC}, 
	country={USA}
}

\affiliation[HUT]{
	organization={College of Electrical Engineering}, 
	addressline={Henan University of Technology}, 
	city={Zhengzhou}, 
	postcode={450001}, 
	state={Henan}, 
	country={China}
}

\cortext[co]{Sangeetha Balabhadra and Haoming Xu are co-first authors and contributed equally to this work.}
\cortext[cor]{Corresponding authors.}


\begin{abstract}
	The formation of heterogeneous lanthanide-ion clusters in CaF$_2$ was investigated experimentally and computationally. CaF$_2$ nanoparticles co-doped with 20~mol\% Yb$^{3+}$ and 2~mol\% Ln$^{3+}$ (Ln$^{3+}$ = Ce$^{3+}$, Pr$^{3+}$, Nd$^{3+}$, Sm$^{3+}$, Eu$^{3+}$, Gd$^{3+}$, Ho$^{3+}$, Er$^{3+}$, and Tm$^{3+}$) were synthesized via a hydrothermal method. The structural and morphological properties were characterized using powder X-ray diffraction, dynamic light scattering, and transmission electron microscopy techniques. High-resolution Fourier transform infra-red spectroscopy revealed the presence of Yb$^{3+}$ isolated cubic centers and various cluster sites. The relative concentration of the clusters varied with the choice of the co-doping ion. Calculations based on density functional theory were used to estimate the formation energies and local coordination structures of different clusters. The calculations indicate that the neutral $C_{4v}$ aggregations containing Ln$^{3+}$ tend to decrease across the lanthanide series, while the negatively charged derivatives of hexameric clusters are relatively constant. 
	This variation matches the experimental results. This study advances understanding of the clustering mechanisms in lanthanide-doped CaF$_2$ nanoparticles and has implications for luminescence optimization in advanced nanomaterials.
\end{abstract}


\end{frontmatter}

\section{Introduction}

Georges Boulon made many significant contributions to spectroscopy. Among these was his work on laser and upconversion  materials containing Yb$^{3+}$. One of the hosts he studied extensively was calcium fluoride (CaF$_2$), which has many desirable qualities, including a large bandgap ($\sim$12~eV), low phonon energy ($\sim$495~cm$^{-1}$) and chemical stability. It is optically transparent in a range from UV to NIR. Heterogeneous trivalent lanthanide ion (Ln$^{3+}$) clusters in alkaline earth fluorides have been studied since the 1970s due to the potential of clustering to affect the energy transfer process and therefore on the luminescence properties of the material~\cite{PhysRevB.20.4308}. Crystals of alkaline earth fluorides have a simple cubic fluorite structure with the fluorine anions (F$^-$) arranged in a simple cubic array and alkaline-earth cations such as calcium (Ca$^{2+}$) in half of the cubic sites of the structure~\cite{10.1119/1.1973685}. These crystals are easily activated by trivalent lanthanide ions (Ln$^{3+}$) owing to the similar ionic radius to that of Ca$^{2+}$ ions~\cite{C4RA08099H}. 

At low activator concentrations ($\sim$0.01~mol\% or less), isolated  Ln$^{3+}$ centers with local and non-local charge compensation dominate. As the concentration increases beyond $\sim$0.05~mol\%, cluster formation (i.e., defect aggregation) begins to occur~\cite{Catlow1984}. 
We adopt the well-established Kr\"{o}ger-Vink notation $\mathrm{D}_\mathrm{S}^C$ to describe point defects~\cite{Guth2014}, where element $\mathrm{D}$ occupies a site in the host occupied originally by element $\mathrm{S}$, with an effective net charge $C$ relative to the perfect lattice ($\prime,~\times,~\bullet$ being $-1$, $0$ and $+1$, respectively). In particular, a vacancy is represented by setting the element $\mathrm{D}$ as $V$, while an interstitial is represented by setting the subscript $\mathrm{S}$ as $i$. 
Cluster centers may occur as dimers (negatively charged $\bigl[2\mathrm{Ln}_\mathrm{Ca}{:}3\mathrm{F}_i\bigr]^\prime$), trimers (negatively charged $\bigl[3\mathrm{Ln}_\mathrm{Ca}{:}4\mathrm{F}_i\bigr]^\prime$)~\cite{10.1063/1.117959}, and even larger, more complex clusters, including hexamers (positively charged $\bigl[6\mathrm{Ln}_\mathrm{Ca}{:}8V_\mathrm{F}{:}13\mathrm{F}_i\bigr]^\bullet$) and their variants~\cite{PhysRevB.72.014127, Nikiforov2005, PhysRevB.90.125124, PhysRevB.78.085131}. 

Our previous experimental studies on CaF$_2$ singly doped with varying concentrations of Yb$^{3+}$ ions have shown that clustering in nanocrystals is very different from clustering in bulk crystals~\cite{BALABHADRA2020155165, BALABHADRA2021110736}. In bulk crystals grown using the high-temperature vertical Bridgman technique, even at low doping concentrations (0.1~mol\%), the absorption spectra exhibit features corresponding to Yb$^{3+}$ clusters, in addition to isolated Yb$^{3+}$ ion centers such as $O_h$, $C_{3v}$, and $C_{4v}$ centers. However, in nanoparticles prepared using hydrothermal method at the same doping concentration, the cluster peaks are undetectable, and the absorption spectra are dominated by isolated cubic $O_h$ point defects.
As the Yb$^{3+}$ concentration increases, the cluster-related absorption peaks emerge in nanoparticles. Nevertheless, the relative intensity of the cluster peaks in nanoparticles are consistently lower than that in bulk crystals at the  same doping concentrations.
This, and other work, indicates that several parameters such as dopant, temperature, and preparation process can affect the number of ions that occupy the various sites in this material, resulting in potentially tunable spectroscopic properties. 

Various techniques namely NMR~\cite{PhysRevB.21.1627}, Raman~\cite{PhysRevB.78.085131}, EXAFS~\cite{Catlow1984}, EPR~\cite{Falin2004, PhysRev.155.279, 10.1098/rspa.1969.0017}, X-ray and neutron diffraction~\cite{LAVAL19881300} and luminescence spectroscopy~\cite{PhysRevB.78.085131, Catlow_1973, WELLS1997977} have been employed for Ln$^{3+}$ site-distribution measurements. In contrast to the well-studied bulk materials~\cite{KALLEL2014261, Kaczmarek2003, 10.1063/1.443882} (mainly single crystals and ceramics), reports on the site distribution of Ln$^{3+}$ in nanoscale alkaline earth fluorides remain limited~\cite{PIERPOINT2020117584, C4TC01214C}. Hitherto, the lanthanide ion doped alkaline earth fluorides have been used in a wide variety of applications in bioimaging~\cite{10.1021/nn202490m, DENG2015154}, thermometry~\cite{thno38091, CAI201613990}, lighting~\cite{YE20112282}, and drug delivery~\cite{10.1002/chem.200903137}. Several strategies have been explored to enhance luminescent properties, such as different dopant, concentration of dopant and foreign dopant effect (Mg$^+$, Li$^+$). The local structural environment in the heterogeneous ions in these nanoparticles upon doping of Ln$^{3+}$ is crucial to understand the energy transfer process and, consequently, the resulting luminescence properties. 

The present work reports Yb$^{3+}$/Ln$^{3+}$ (Ln$^{3+}$ = Ce$^{3+}$, Pr$^{3+}$, Nd$^{3+}$, Sm$^{3+}$, Eu$^{3+}$, Gd$^{3+}$, Ho$^{3+}$, Er$^{3+}$, and Tm$^{3+}$) co-doped CaF$_2$ nanoparticles. The as-synthesized nanoparticle crystal structure and hydrodynamic sizes have been identified by powder X-ray diffraction (PXRD) and dynamic light scattering (DLS) techniques. The absorption spectra of as-synthesized nanoparticles were performed using a high-resolution Fourier transform infra-red (FTIR) spectrometer. The effect of type of Ln$^{3+}$ in the series on the Yb$^{3+}$ cluster formation is systematically investigated. In addition, DFT calculations were performed to obtain the local structure and formation energies of various possible clusters. First-principles calculations based on the density functional theory (DFT) have been shown to yield reliable information about not only the energy but also the atomic coordinate of the supercell~\cite{VandeWalle_1.1682673, VandeWalle_RevModPhys.86.253}. Based on the formation energies, we can estimate the ratios of different clusters.

\section{Methodology}

\subsection{Sample synthesis}

20~mol\% Yb$^{3+}$ ion doped and 20~mol\% Yb$^{3+}$/ 2~mol\% Ln$^{3+}$ ion co-doped CaF$_2$ upconverting nanoparticles were prepared by a hydrothermal method at 190~$^\circ$C for 6 hours using sodium citrate as capping agent, as described in Ref.~\cite{BALABHADRA2020155165}. 

\subsection{Experimental characterization}

\subsubsection{Powder X-Ray Diffraction}

Phase identification of the as-synthesized samples was inferred from their X-ray diffraction patterns.  Corresponding diffractograms were collected on a Rigaku SmartLab powder X-ray diffractometer with CuK$\alpha$1 radiation, 1.5406~\r{A}, operating at 40~kV and 30~mA, in the 2$\theta$ range 20$^\circ$ to 90$^\circ$ with a 0.01$^\circ$ step size in the reflection scanning mode. The reference data were taken from the International Centre for Diffraction Data (ICDD) database. 

\subsubsection{Dynamic Light Scattering}

Dynamic light scattering (DLS) measurements were carried out using a Malvern Zetasizer Nano ZS instrument operating at 532~nm with a 50~mW laser. The data were acquired for sodium citrate capped CaF$_2$:Yb$^{3+}$ nanoparticles dispersed to have 0.25~wt\% aqueous suspension. The reported values are the average of three measurements.

\subsubsection{Transmission Electron Microscopy (TEM)}

The morphology of the nanoparticles was recorded on a PHILIPS CM200 transmission electron microscope operated at 200~kV having a Gatan Orus CCD camera. The sample for analysis was prepared by placing a droplet of the nanoparticle suspension in acetone on a carbon-coated 300-mesh copper grid, which was then allowed to evaporate overnight in the air at room temperature.

\subsubsection{Fourier-Transform Infrared Spectroscopy}

Temperature dependent infrared absorption measurements were performed using a Bruker Vertex 80 FTIR having 0.075~cm$^{-1}$ resolution with an optical path purged by N$_2$ gas. For the absorption measurements, the air-dried powders of nanoparticles were pressed using a pellet maker in order to form a thin pellet ($\sim$1~mm) which was then placed into a small copper sample holder. The samples were cooled in closed-cycle helium cryostat with temperature variability (10--298~K) provided by a resistive heating element.

\subsection{Computational details}

The Vienna Ab-initio Simulation Package (VASP)~\cite{VASP_PhysRevB.47.558, VASP_PhysRevB.49.14251} was employed for the calculations.
The projector augmented-wave (PAW) pseudo-potentials~\cite{PAW_PhysRevB.50.17953, PAW_PhysRevB.59.1758} were used to treat the interactions between valence electrons and ion cores. The Ca:3p$^6$4s$^2$, F:2s$^2$2p$^5$, Ce, Pr, Nd, Pm, Sm:5s$^2$5p$^6$5d$^1$6s$^2$ and Eu, Gd, Tb, Dy, Ho, Er, Tm, Yb:5p$^6$5d$^1$6s$^2$ electrons were treated as the valence electrons. The Perdew-Burke-Ernzerhof (PBE) exchange-correlation functional within generalized gradient approximation (GGA)~\cite{GGA_PhysRevLett.77.3865} was used. The ionic positions were fully relaxed until the residual force acting on each ion is less than 0.01~eV/\r{A}. 
The lattice parameter was obtained by setting the energy cut-off as 520~eV in the plane-wave basis set and using large $9\times9\times9$ $k$-points grids for accuracy. Then a supercell composed of $3\times3\times3$ of the unit cell with the obtained lattice parameter was constructed to accommodate the defects. 
The total energies were calculated for the pristine supercell and supercells containing individually a variety of potentially energetic favourable complexes formed with tripositive lanthanide ions at calcium sites together with  monovalent anionic interstitials. For computational efficiency, these supercell calculations were performed using a single $\Gamma$-point and an energy cut-off of 400~eV, with the lattice constant fixed. Charge corrections, including both the image-charge interaction
correction ($E_\text{IIC}$) and the potential alignment correction ($q\Delta V_\text{NAP}$), were included for the total energies of charged defects~\cite{correction_1.5029818}.

\subsubsection{Formation energy of defect}

The formation energy of a defect $X$ in the charge state of $q$ can be derived from~\cite{VandeWalle_RevModPhys.86.253}:
\begin{equation}
\label{eq:EF0}
\begin{aligned}
	E^f\bigl(X^q,E_F\bigr)=&\bigl(E_\text{tot}\bigl[X^q\bigr]+E_\text{corr}\bigl[X^q\bigr]\bigr)-E_\text{tot}\bigl[\text{bulk}\bigr]\\
	&-\sum_{i}{n_i\mu_i}+q E_F\\
	=&\Delta E_\text{tot}\bigl[X^q\bigr] -\sum_{i}{n_i\mu_i}+q E_F,
\end{aligned}
\end{equation}
where $E_\text{tot}$ is the calculated total energy of the relaxed doped and undoped supercells; $E_\text{corr}$ is the energy corrections~\cite{correction_1.5029818} in the DFT calculations; $n_i$ is the number of the atoms of elements $i$ which are added to ($n_i>0$) and/or removed from ($n_i<0$) the perfect supercell; $\mu_i$ is corresponding atomic chemical potential of the element $i$; $q$ is the charge state of the defect $X$; and $E_F$ is the chemical potential of electrons, i.e., the Fermi energy. $\Delta E_\text{tot}\bigl[X^q\bigr] = \bigl(E_\text{tot}\bigl[X^q\bigr]+E_\text{corr}\bigl[X^q\bigr]\bigr)-E_\text{tot}\bigl[\text{bulk}\bigr]$ is obtained from the DFT calculations based on the supercell method.

\begin{figure*}[tp]
	\centering
        \subfloat[PXRD]{\label{PXRD} \includegraphics[width=0.33\linewidth]{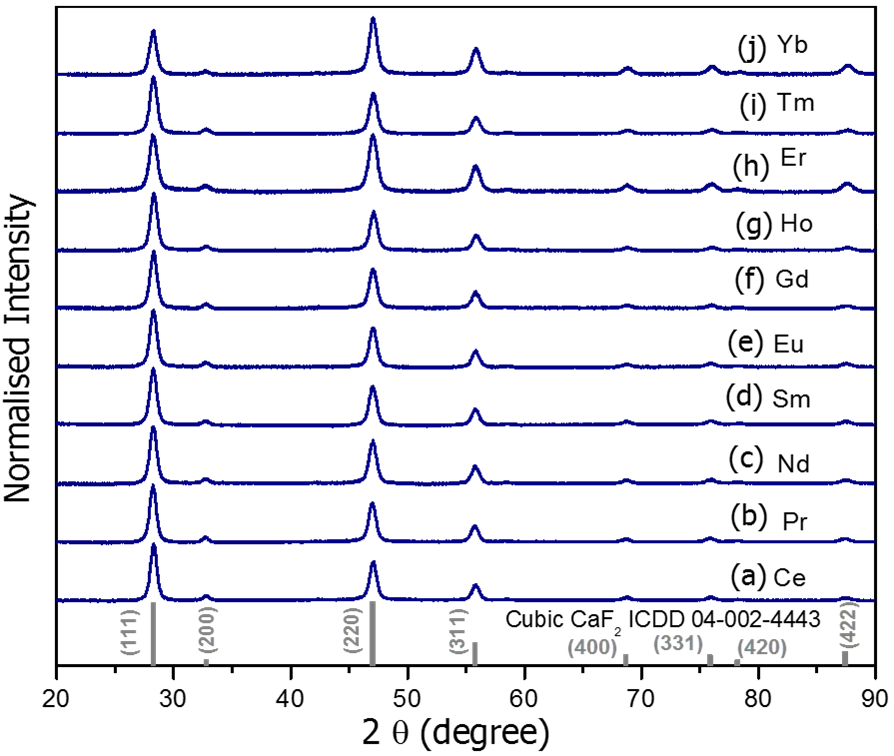}}
	\hfill
	\subfloat[DLS]{\label{DLS} \includegraphics[width=0.35\linewidth]{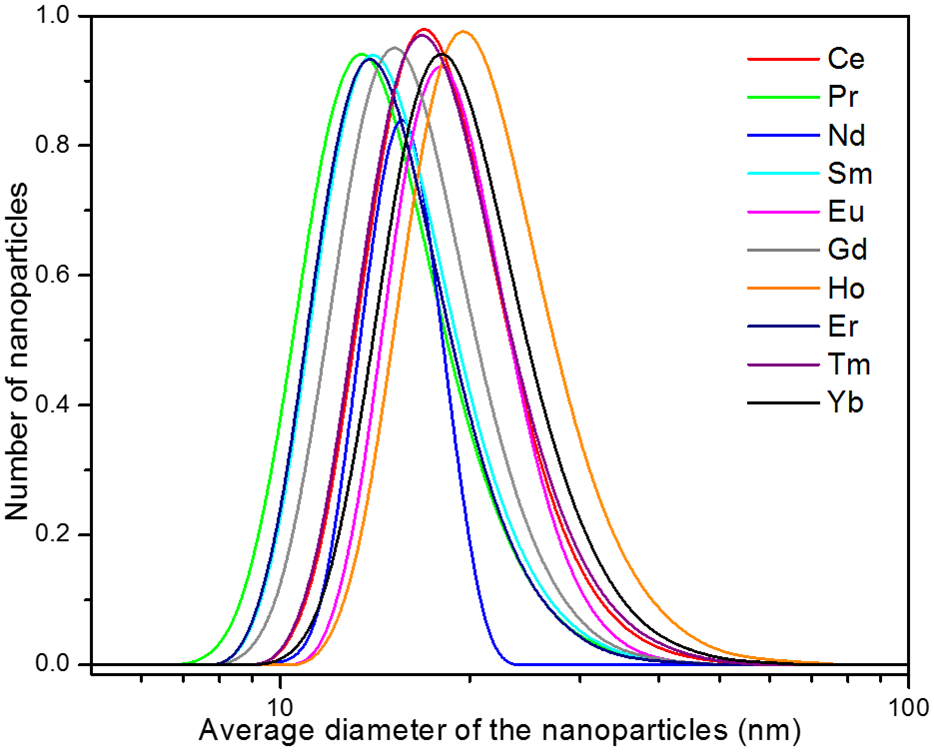}}
	\hfill
	\subfloat[TEM]{\label{TEM} \includegraphics[width=0.3\linewidth]{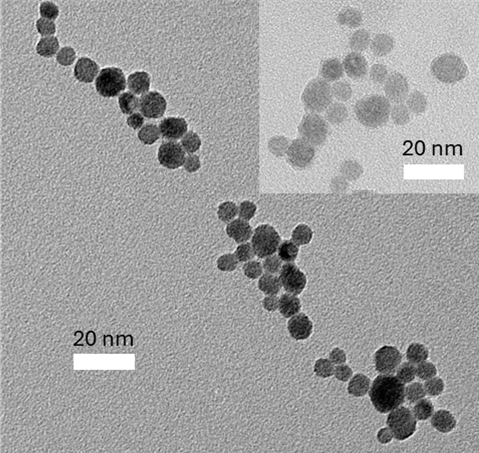}}
	\caption{(a) Powder X-ray diffraction patterns of CaF$_2$ nanoparticles doped with 20~mol\% Yb$^{3+}$ and 2~mol\% Ln$^{3+}$ nanoparticles. The diffraction patterns of cubic CaF$_2$ (ICDD Card No 04-002-4443) are also depicted. (b) Hydrodynamic sizes measured in dynamic light scattering of the nanoparticles. (c) TEM images of CaF$_2$: 20~mol\% Yb$^{3+}$/ 2~mol\% Ho$^{3+}$ nanoparticles. }
	\label{exp}
\end{figure*}

In our calculation of CaF$_2$:Yb$^{3+}$/Ln$^{3+}$ (Ln = Pr--Tm), the valence states of all elements are fixed, allowing electrons to be combined with atoms to form ions of well-defined charge. Eq.~\eqref{eq:EF0} may be rewritten as:
\begin{equation}
\label{eq:EF}
E^f\bigl(X^q\bigr)=\Delta E_\text{tot}\bigl[X^q\bigr]-\sum_{i}{n_i\mu_i},
\end{equation}
where the index $i$ represents  the composition ions Yb$^{3+}$, Ln$^{3+}$, F$^-$ and  Ca$^{2+}$, $n_i$ is the change of the number of the ion $i$ added to ($n_i>0$) and or removed from ($n_i<0$) the pristine supercell, and $\mu_i$ is corresponding chemical potential of these ions. For example, $\mu\bigl(\mathrm{Ln}^{3+}\bigr)=\mu\bigl(\mathrm{Ln}\bigr)-3 E_F$. The $\mu_i$ in Eq.~\eqref{eq:EF} are variable and have to meet the condition of thermal stability for CaF$_2$ host:
\begin{equation}
\label{eq:CaF2}
\mu\bigl(\mathrm{Ca}^{2+}\bigr)+2\mu\bigl(\mathrm{F}^-\bigr)=\mu\bigl(\mathrm{CaF}_2\bigr),
\end{equation}
where the $\mu\bigl(\mathrm{CaF}_2\bigr)$ is the DFT-calculated total energy per formula unit for CaF$_2$ host.

\subsubsection{Proportion of defect clusters}

The concentration $c_i$, expressed as the fraction of calcium sites occupied by  defect $i$, is given approximately by:
\begin{equation}
\label{eq:c}
c_i = g_i \exp\bigl(-\frac{E^f_i}{k_\mathrm{B}T}\bigr),
\end{equation}
where $g_i$ is the degeneracy factor of the defect $i$, $E^f_i$ is the formation energy of the defect $i$ as given by Eq.~\eqref{eq:EF}, $k_\mathrm{B}$ is the Boltzmann constant, and $T$ is the temperature at which the defect stabilizes concentration equilibrates. 

The conditions of charge neutrality and fixed dopant concentrations impose the following constraints:
\begin{align}
    \sum_{i}{q_i~c_i}&=0, \label{eq:q} \\
    \sum_{i}{n^\mathrm{Yb}_i~c_i}&=c^\mathrm{Yb}, \label{eq:Yb} \\
	\sum_{i}{n^\mathrm{Ln}_i~c_i}&=c^\mathrm{Ln}, \label{eq:Ln}
\end{align}
where $q_i$ and $n^\mathrm{Ln}_i$ (with Ln including Yb) denote the net charge and number of Ln atoms in defect cluster $i$, respectively, and $c^\mathrm{Ln}$ (aging including Yb) is the dopant concentration of Ln.

Given $\mu\bigl(\mathrm{CaF}_2\bigr)$ from the calculation and $c^\mathrm{Yb}$ and $c^\mathrm{Ln}$ specified in experiment, the unknown $\mu\bigl(\mathrm{F}^-\bigr)$, $\mu\bigl(\mathrm{Ca}^{2+}\bigr)$, $\mu\bigl(\mathrm{Yb}^{3+}\bigr)$ and $\mu\bigl(\mathrm{Ln}^{3+}\bigr)$ can be determined by solving equations \eqref{eq:CaF2} and \eqref{eq:q}--\eqref{eq:Ln}. The concentration of each defect cluster $c_i$ can then be obtained from Eq.~\eqref{eq:c}. In practice, these equations are solved numerically via an iterative method.

\section{Results and Discussion}

\subsection{Characterization of samples}

Figure~\subref*{PXRD} displays the powder X-ray diffraction patterns of singly-doped 20~mol\% Yb$^{3+}$ and co-doped 20~mol\% Yb$^{3+}$/2~mol\% Ln$^{3+}$ CaF$_2$ nanoparticles. The samples are of pure cubic crystalline phase (space group F$m\overline{3}m$), in agreement with ICDD card 04-002-4443 for CaF$_2$.
Our DFT calculations, performed using the PBEsol exchange-correlation functional within the GGA framework, yield an optimized cubic lattice constant of $a = 5.413~\text{\r{A}}$. This is in excellent agreement with the experimental value of $a = 5.462~\text{\r{A}}$~\cite{Cheetham_1971}, with a deviation of less than 1\%. This close agreement confirms the reliability of our computational approach for the subsequent cluster simulations. From the diffractograms the average size of the crystallite domains in nanoparticles could be estimated using Scherrer's equation using the full-width half-maximum value of the 2$\theta$ diffraction peak around 28$^\circ$ (\textit{i.e.} the $\left(111\right)$ plane).

\begin{table}[tp]
    \caption{Crystallite and hydrodynamic sizes (nm) obtained from PXRD and DLS measurements for CaF$_2$ nanoparticles co-doped with Ln$^{3+}$/Yb$^{3+}$ (sample nos. 1--9) and singly doped with Yb$^{3+}$ (sample no. 10).}
    \label{sizes}
    \centering
    \begin{tabular}{cccc}
        \hline
        \hline
        No. & Dopant ions & Crystallite & Hydrodynamic \\
        \hline
        1 & Yb$^{3+}$/Ce$^{3+}$ & 12.47 & 16.78 \\
        2 & Yb$^{3+}$/Pr$^{3+}$ & 12.22 & 13.55 \\
        3 & Yb$^{3+}$/Nd$^{3+}$ & 11.60 & 15.60 \\
        4 & Yb$^{3+}$/Sm$^{3+}$ & 11.68 & 13.93 \\
        5 & Yb$^{3+}$/Eu$^{3+}$ & 11.75 & 17.97 \\
        6 & Yb$^{3+}$/Gd$^{3+}$ & 11.67 & 15.06 \\
        7 & Yb$^{3+}$/Ho$^{3+}$ & 11.70 & 19.53 \\
        8 & Yb$^{3+}$/Er$^{3+}$ & 10.51 & 13.90 \\
        9 & Yb$^{3+}$/Tm$^{3+}$ & 11.38 & 16.79 \\
        10 & Yb$^{3+}$ & 11.70 & 18.13 \\
        \hline
        \hline
    \end{tabular}
\end{table}

The hydrodynamic size distribution of the nanoparticles was measured using dynamic light scattering and shown in Fig.~\subref*{DLS}. The hydrodynamic sizes are in good agreement with the average crystallite sizes obtained from the PXRD results (Table~\ref{sizes}). The transmission electron microscopy image of one of the prepared nanoparticles (CaF$_2$:20~mol\% Yb$^{3+}$/2~mol\% Ho$^{3+}$) is shown in Fig.~\subref*{TEM}. The nanoparticles are homogenously distributed and spherical in shape with an average diamter of 13.4~nm (obtained by measuring around 80 nanoparticles) similar to the observed size from PXRD and DLS.

\begin{figure}[tp]
	\centering
	\includegraphics[width=\linewidth]{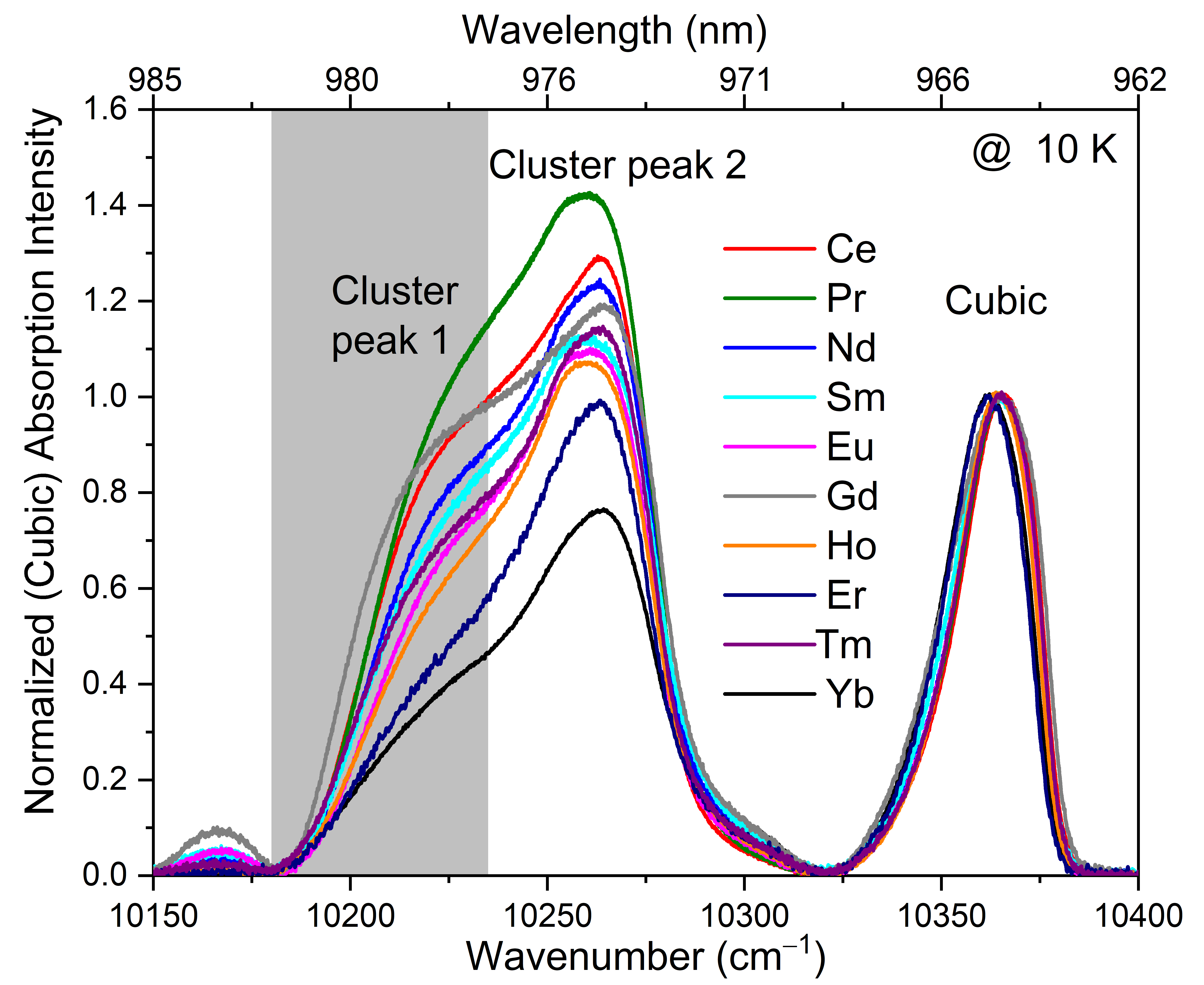}
	\caption{FTIR absorption spectra Yb$^{3+}$ ${}^2\mathrm{F}_{7/2}{\rightarrow}{}^2\mathrm{F}_{5/2}$ inter-multiplet measured at 10~K for singly-doped 20~mol\% Yb$^{3+}$ and co-doped 20~mol\% Yb$^{3+}$/2~mol\% Ln$^{3+}$ CaF$_2$ nanoparticles.}
	\label{FTIR}
\end{figure}

\subsection{FTIR absorption measurements}

Figure~\ref{FTIR} shows the high-resolution Fourier-transform infrared absorption measurements recorded for upconverting nanoparticles at 10~K. The absorption spectra recorded for all of the samples show characteristic absorption bands arising from the Yb$^{3+}$ ${}^2\mathrm{F}_{7/2}{\rightarrow}{}^2\mathrm{F}_{5/2}$ inter-multiplet transition. Comparing absorption depths between samples is not feasible owing to the different thickness and distribution of nanoparticles in the pellet used to measure the absorbance of the nanoparticles, so each spectrum is normalized to the absorbance of the Yb$^{3+}$ cubic site. The co-dopant 2~mol\% Ln$^{3+}$ is varied across the series, while the 20~mol\% concentration of Yb$^{3+}$ is held constant. The absorption spectra are completely dominated by single Yb$^{3+}$ ion cubic centers (10325--10400~cm$^{-1}$) and cluster sites (10150--10320~cm$^{-1}$) as reported in the literature~\cite{PhysRevB.90.125124, PhysRevB.78.085131}. 
The spectral feature associated with the cluster sites appears to be a composite of two features, suggesting contributions from distinct cluster geometries. The low-energy shoulder is designated as ``cluster peak 1'' ($\sim$10225~cm$^{-1}$), while the prominent feature at higher energy is designated as ``cluster peak 2'' ($\sim$10265~cm$^{-1}$).

Because cluster 1 is significantly broader than cluster 2, the observed variation in the peak height of cluster 2 across the lanthanide series is strongly influenced by spectral overlap from cluster 1.
The ratio of the peak areas of cluster peak 1 and cluster peak 2 to the total cubic area (determined by fitting Gaussian functions) in Ln$^{3+}$/Yb$^{3+}$ co-doped and Yb$^{3+}$ singly doped CaF$_2$ nanoparticles is shown in figure~\subref*{ratio}. 
The area of cluster peak 1 changes dramatically with the co-doped Ln$^{3+}$, whereas the area of cluster peak 2 is relatively constant. This is a surprising result, since the co-doped ions have a much lower concentration than Yb$^{3+}$.

\begin{figure}[t]
	\centering
	\subfloat[\texorpdfstring{$O_h$ $\mathrm{Ln}_\mathrm{Ca}^\bullet$}{Oh Ln}]{\label{Oh} \includegraphics[width=0.2\linewidth]{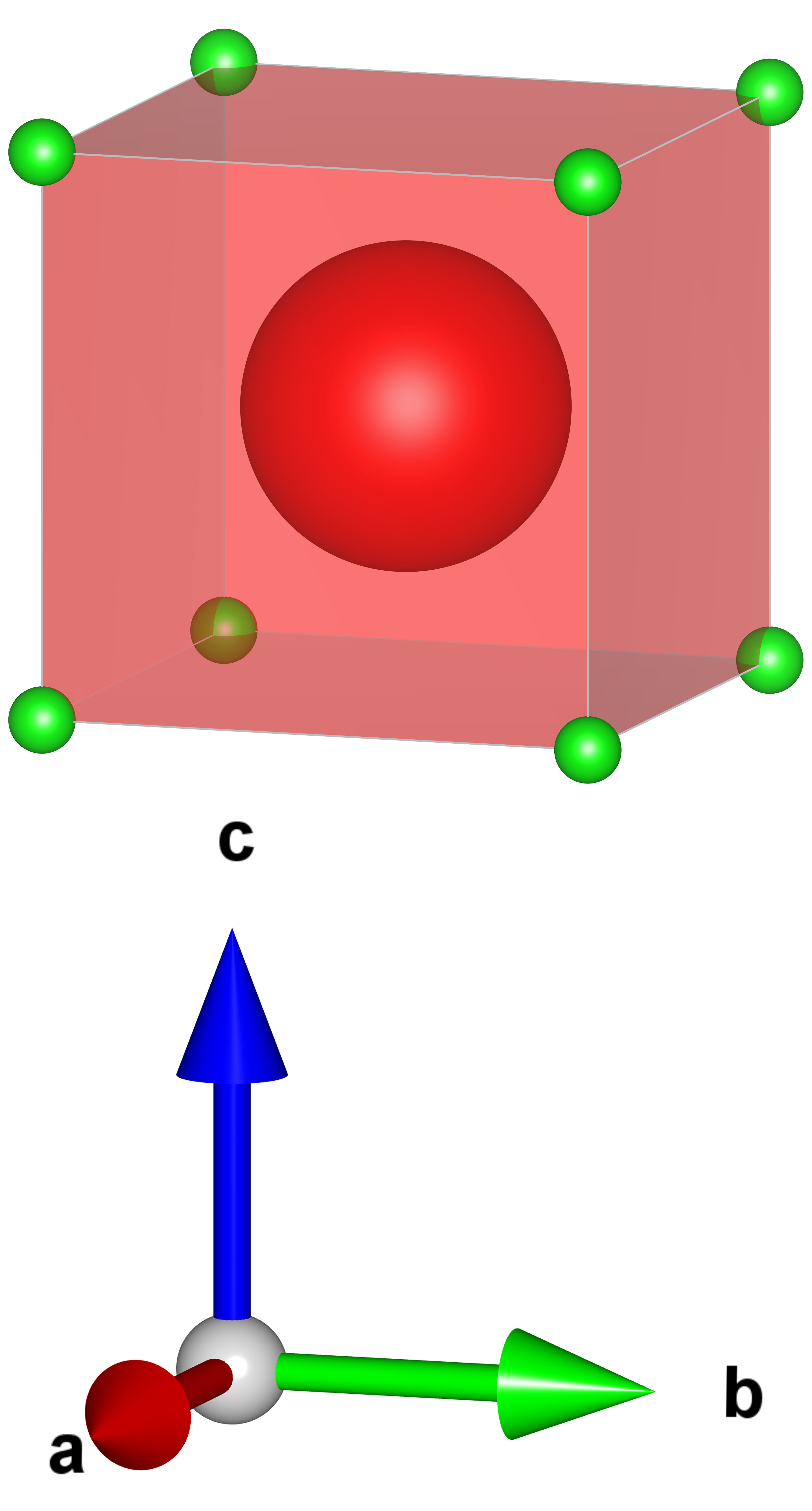}}
	\subfloat[\texorpdfstring{$C_{3v}$ $\bigl[\mathrm{Ln}_\mathrm{Ca}{:}\mathrm{F}_i\bigr]^\times$}{C3v Ln:F}]{\label{C3v} \includegraphics[width=0.4\linewidth]{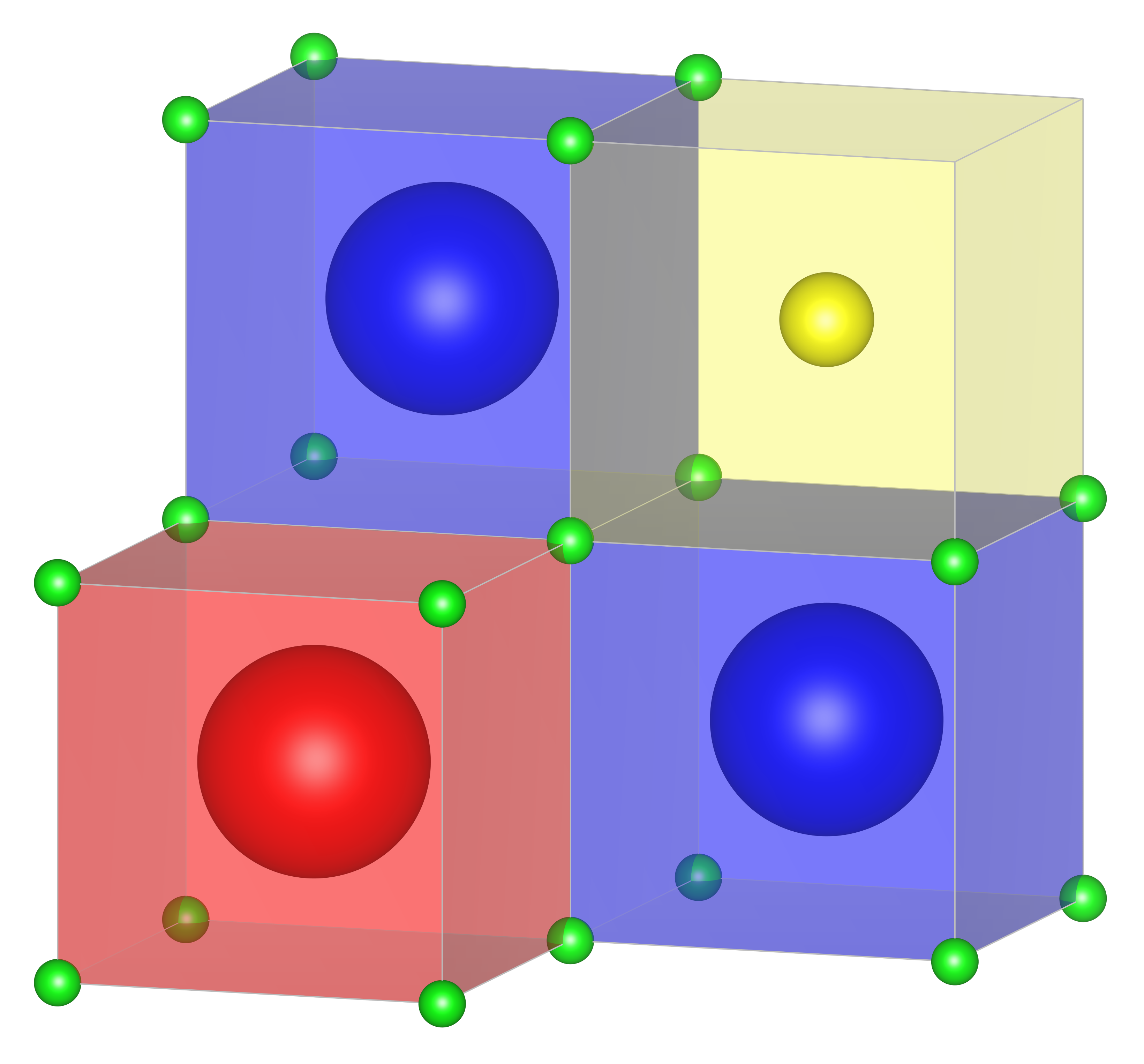}}
	\subfloat[\texorpdfstring{$C_{4v}$ $\bigl[\mathrm{Ln}_\mathrm{Ca}{:}\mathrm{F}_i\bigr]^\times$}{C4v Ln:F}]{\label{C4v}\includegraphics[width=0.4\linewidth]{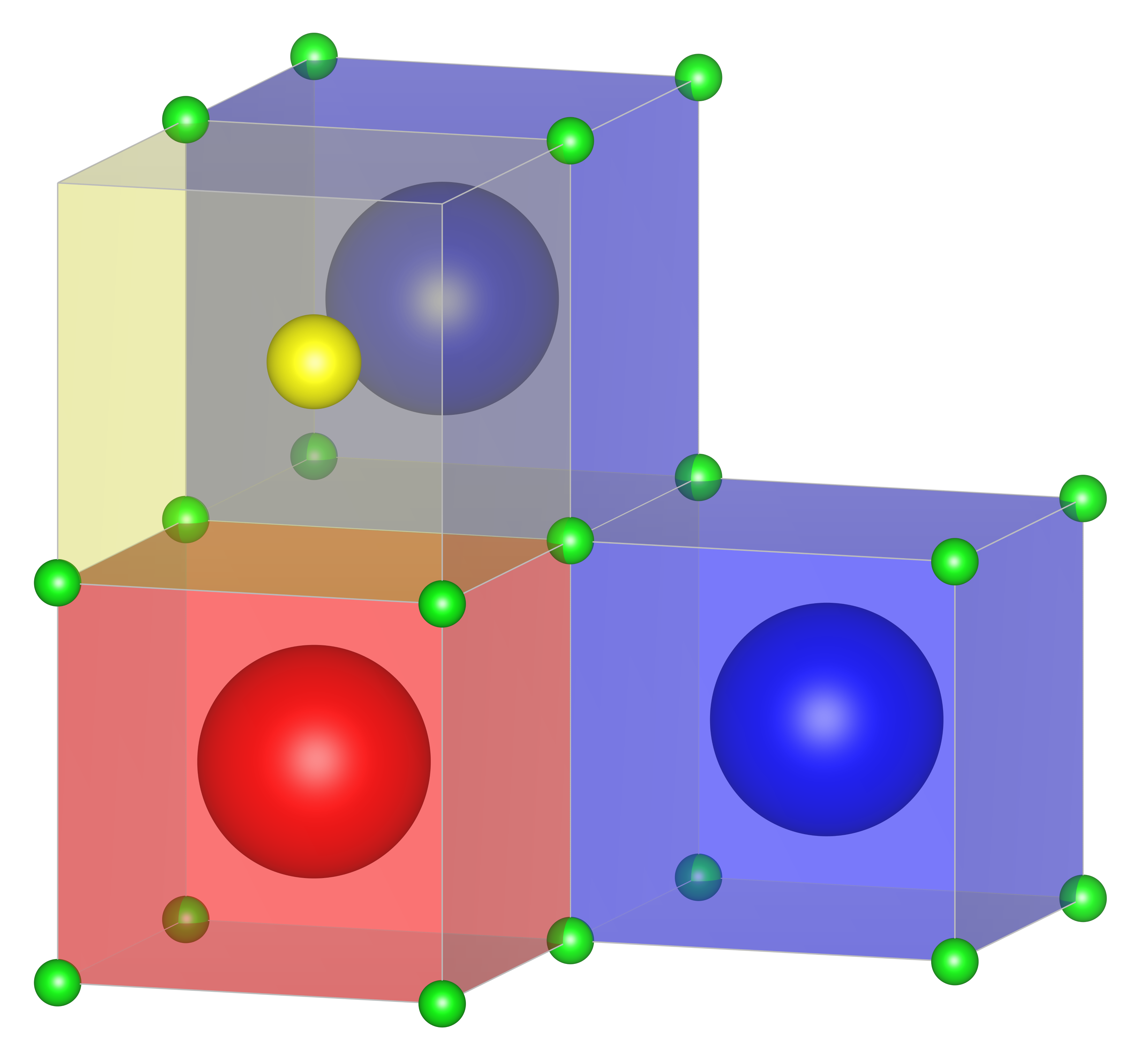}}
	\hfill
	\subfloat[Unrelaxed \texorpdfstring{$\bigl[3\mathrm{Ln}_\mathrm{Ca}{:}3\mathrm{F}_i\bigr]^\times$}{3Ln:3F}]{\label{host33}\includegraphics[width=0.5\linewidth]{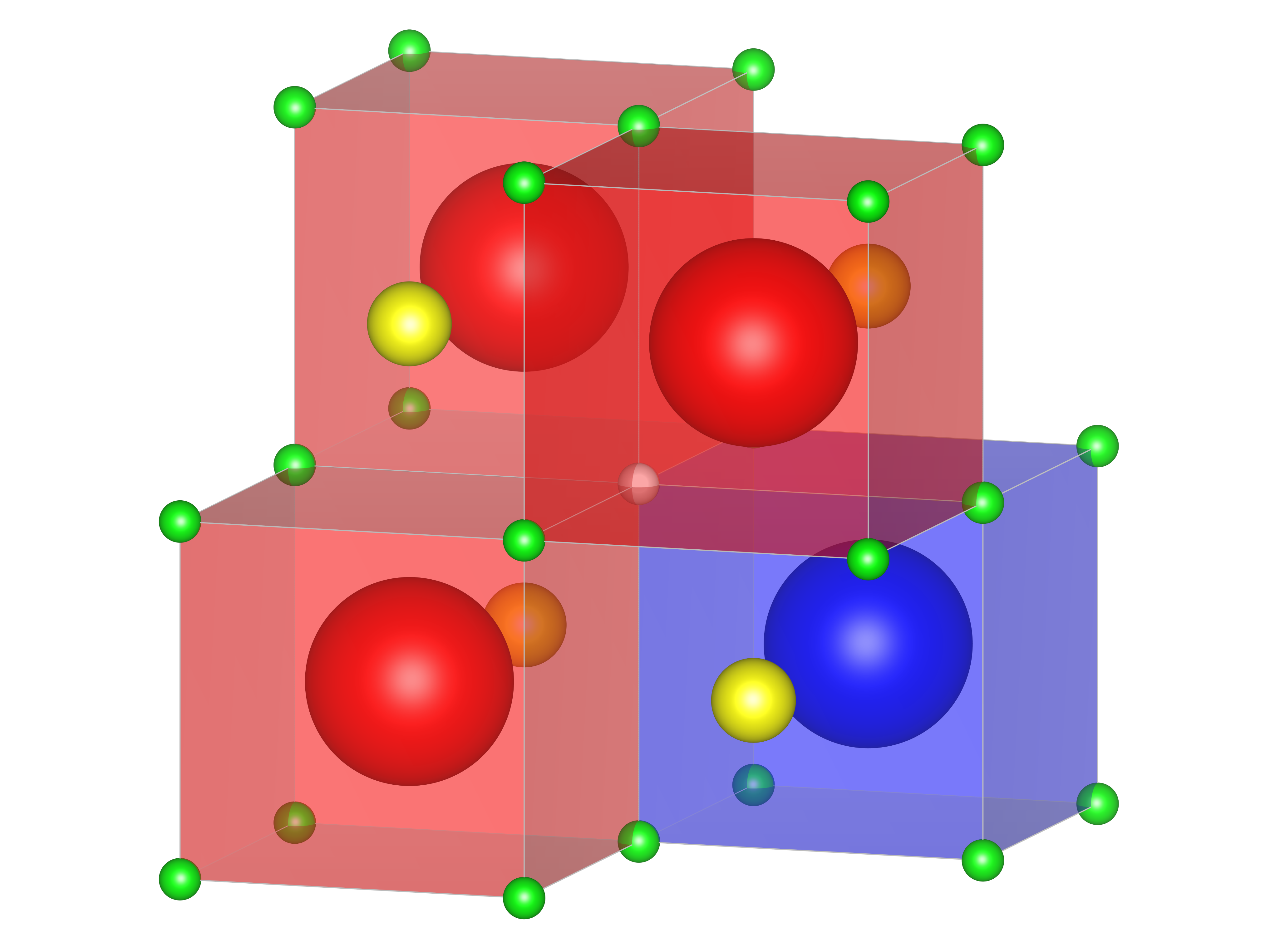}}
	\subfloat[Relaxed \texorpdfstring{$\bigl[3\mathrm{Ln}_\mathrm{Ca}{:}3\mathrm{F}_i\bigr]^\times$}{3Ln:3F}]{\label{cluster33}\includegraphics[width=0.5\linewidth]{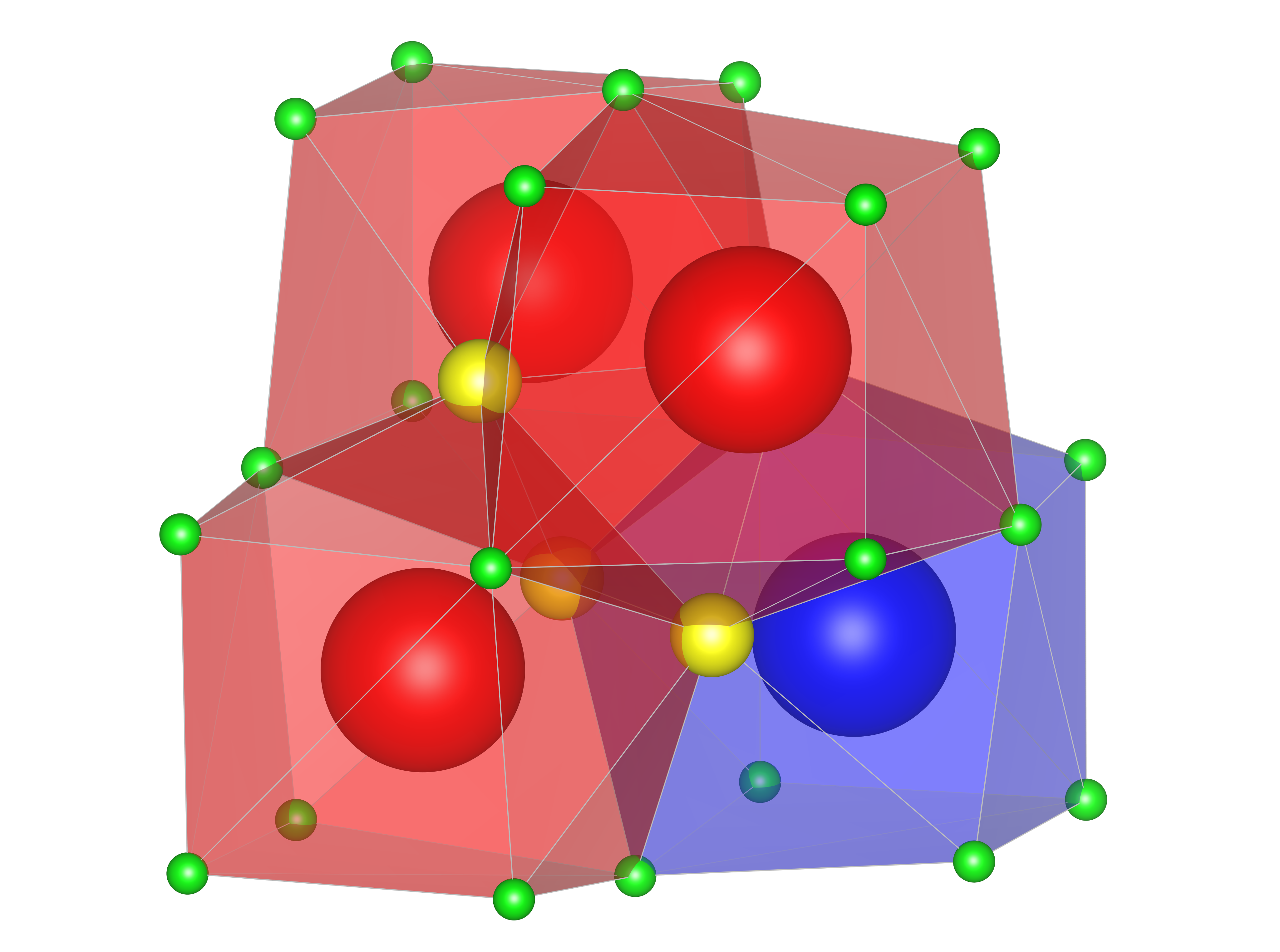}}
	\hfill
	\subfloat[Unrelaxed \texorpdfstring{$\bigl[3\mathrm{Ln}_\mathrm{Ca}{:}4\mathrm{F}_i\bigr]^\prime$}{1Ln:2Ln:4F}]{\label{host34}\includegraphics[width=0.5\linewidth]{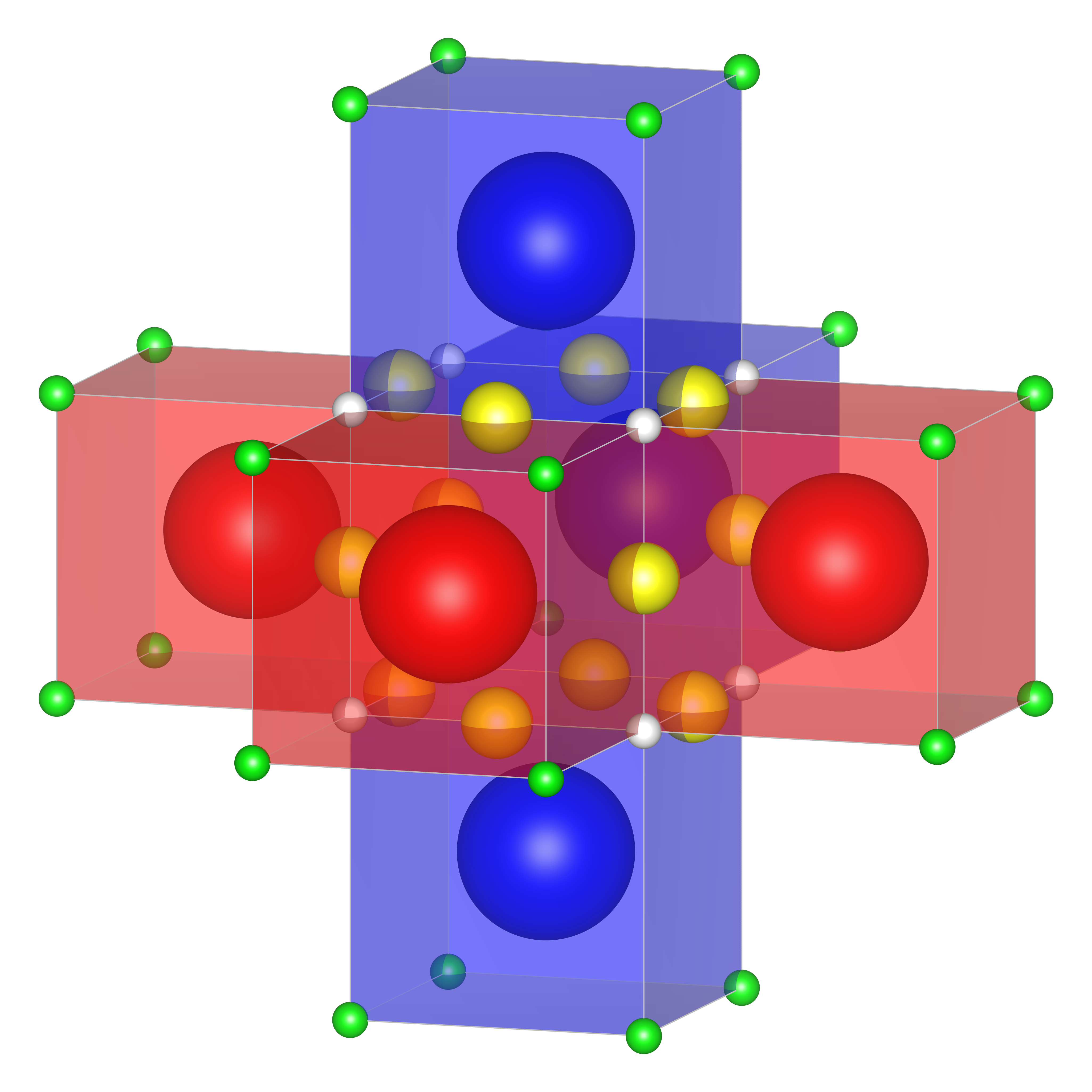}}
	\subfloat[Relaxed \texorpdfstring{$\bigl[3\mathrm{Ln}_\mathrm{Ca}{:}4\mathrm{F}_i\bigr]^\prime$}{1Ln:2Ln:4F}]{\label{cluster34}\includegraphics[width=0.5\linewidth]{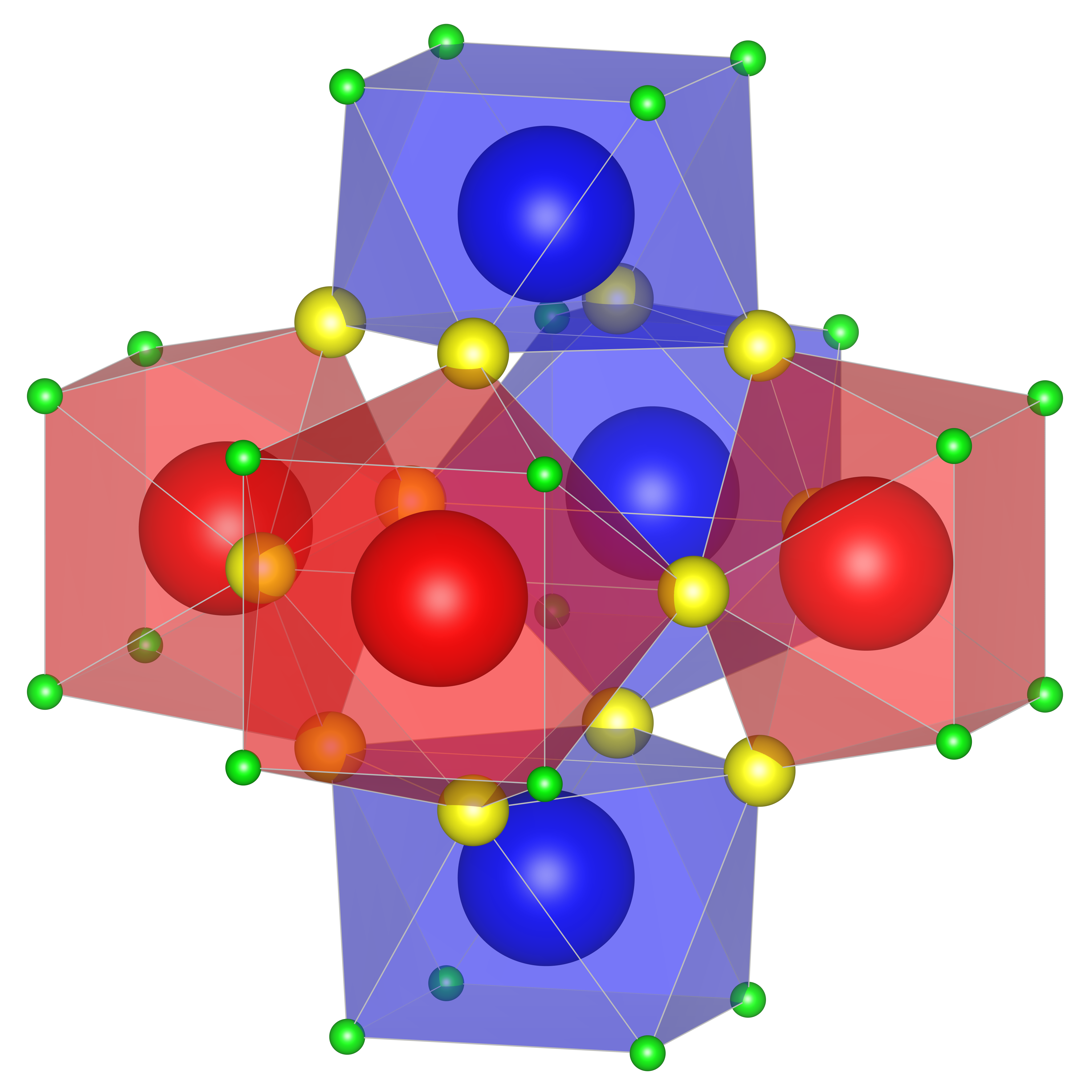}}
	\caption{Coordination structure of (a) $O_h$ $\mathrm{Ln}_\mathrm{Ca}^\bullet$, (b) $C_{3v}$ $\bigl[\mathrm{Ln}_\mathrm{Ca}{:}\mathrm{F}_i\bigr]^\times$, (c) $C_{4v}$ $\bigl[\mathrm{Ln}_\mathrm{Ca}{:}\mathrm{F}_i\bigr]^\times$, (d, e) $\bigl[3\mathrm{Ln}_\mathrm{Ca}{:}3\mathrm{F}_i\bigr]^\times$ and (f, g) $\bigl[3\mathrm{Ln}_\mathrm{Ca}{:}4\mathrm{F}_i\bigr]^\prime$ clusters in alkaline earth fluoride system. The legend is as follows: blue: Ca$^{2+}$ ion, green: F$^-$ ion, red: Ln$_\mathrm{Ca}^\bullet$ substitutional, yellow: F$_i^\prime$ interstitial, white: $V_\mathrm{F}^\bullet$ vacancy.}
	\label{structure}
\end{figure}

\subsection{Calculated coordination structure of clusters}

We now use DFT calculations to examine the relative stability of co-doped clusters. When doped with Ln$^{3+}$, the Ca$^{2+}$ cation replaced by Ln$^{3+}$ to form Ln${_\mathrm{Ca}^\bullet}$ substitutional, with subsequent compensation of the excessive positive charge potentially leading to defects such as interstitial fluorine (F$_i^\prime$) modifying the cubic ($O_h$) symmetry. In the $O_h$ centers, 
the Ln$^{3+}$ substitutes for Ca$^{2+}$ without local charge compensation (Fig.~\subref*{Oh}). The two other site symmetries (monomers) are formed with a single pair~\cite{PhysRevB.41.10799, C9TC04786G}. In the tetragonal ($C_{4v}$) symmetry, one Ln${_\mathrm{Ca}^\bullet}$ is compensated in a near-neighbor position by a fluoride interstitial ion (F$_i^\prime$) in the $\langle001\rangle$ direction (Fig.~\subref*{C4v}), while in the trigonal ($C_{3v}$) symmetry the Ln${_\mathrm{Ca}^\bullet}$ is compensated by a next-nearest-neighbor F$_i^\prime$ in the $\langle111\rangle$ direction (Fig.~\subref*{C3v})~\cite{CIRILLOPENN1991505}. 

There are various heterogeneous clusters, and here we introduce two that are dominant in the DFT results. The $\bigl[3\mathrm{Ln}_\mathrm{Ca}{:}3\mathrm{F}_i\bigr]^\times$ cluster is formed by three Ln${_\mathrm{Ca}^\bullet}$ substitutional and three near-neighbor F$_i^\prime$ interstitial ions. This is similar to several $C_{4v}$ monomers aggregating together (Fig.~\subref*{host33}). As the $C_{4v}$ $\bigl[\mathrm{Ln}_\mathrm{Ca}{:}\mathrm{F}_i\bigr]^\times$ is neutral, it could cluster in any number without the need for additional charge compensation. 
The $\bigl[3\mathrm{Ln}_\mathrm{Ca}{:}4\mathrm{F}_i\bigr]^\prime$ (Fig.~\subref*{host34}) cluster is a derivative of hexameric $\bigl[6\mathrm{Ln}_\mathrm{Ca}{:}8V_\mathrm{F}{:}13\mathrm{F}_i\bigr]^\bullet$~\cite{PhysRevB.72.014127}. 
This hexameric cluster can have a variety of different clusters through changes in the number of $\mathrm{Ln}_\mathrm{Ca}^\bullet$ substitutionals and the central $\mathrm{F}_i^\prime$ interstitial. In addition, the Ln$^{3+}$ ions have a similar coordination environment in the hexameric cluster to that in the trifluoride~\cite{CAI2022119058}.

\begin{figure}[tb]
	\centering
	\subfloat[Ratio of absorption peak areas]{\label{ratio}\includegraphics[width=\linewidth]{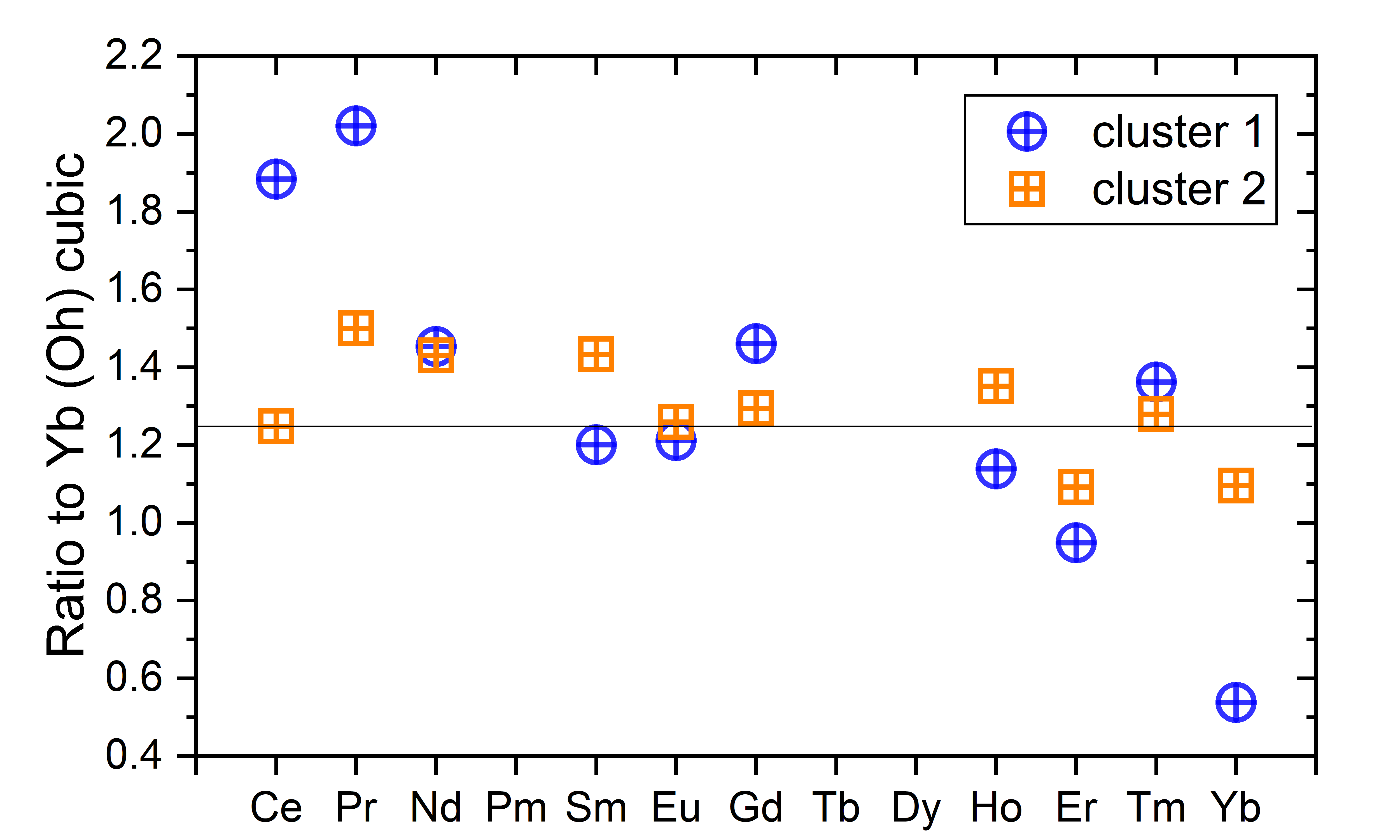}}
	\hfill
	\subfloat[Calculated concentration of clusters]{\label{concentration}\includegraphics[width=\linewidth]{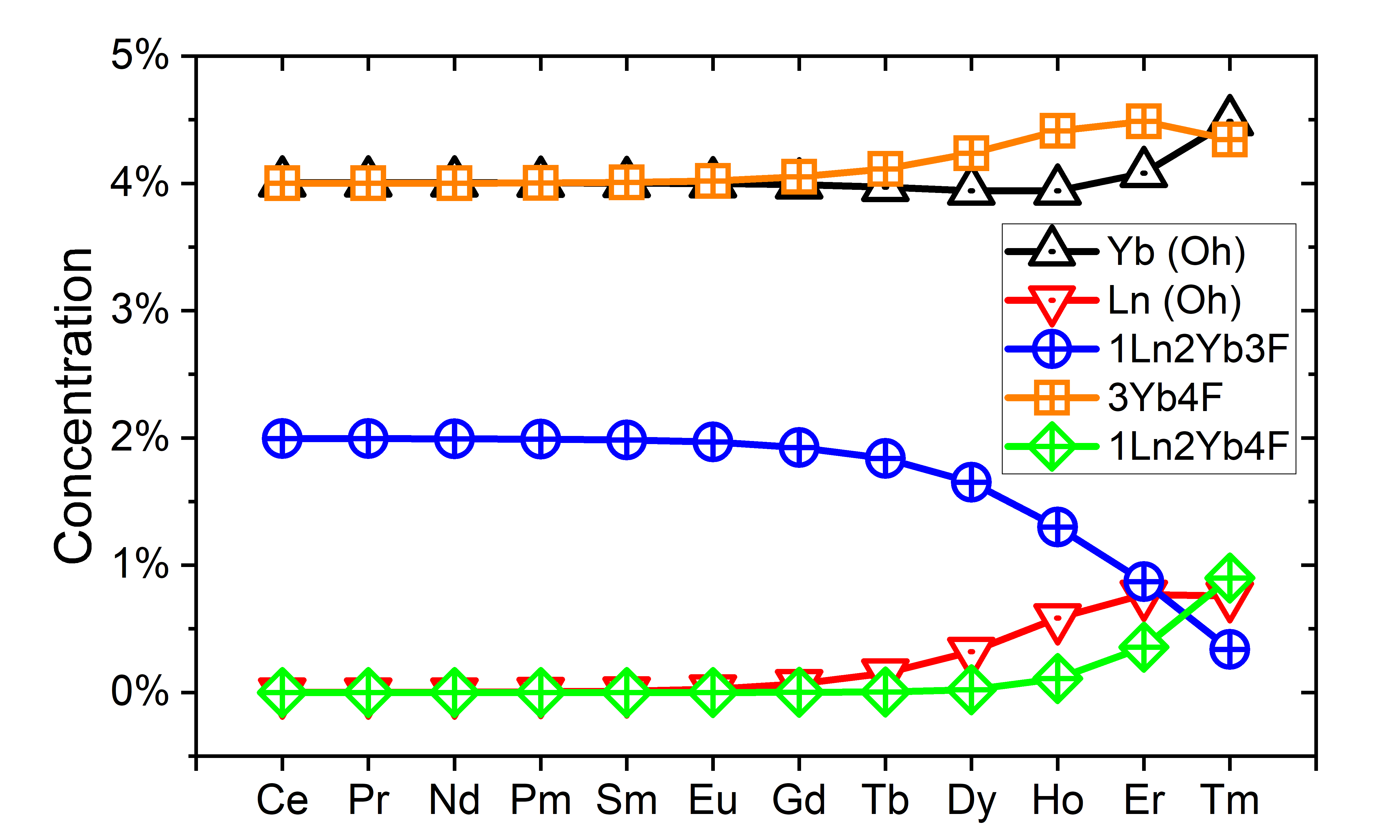}}
	\caption{Behaviour of heterogeneous clusters across the lanthanide series in Ln$^{3+}$/Yb$^{3+}$ doped CaF$_2$. (a) Ratio of the peak areas obtained from experimental FTIR absorption spectra. (b) Concentration of dominant clusters obtained from the DFT calculations. Other clusters contain up to three Ln and Yb ions combined have concentrations far too low to be shown on this scale.}
	\label{cluster}
	
\end{figure}

  The formation energy of various clusters (as well as point defects) under Yb singly doped CaF$_2$ crystals are listed in Table S1 in Supplementary Material, and the chemical potentials and formation energies of the clusters for different co-doping lanthanides are listed in Table S2 in Supplementary Material.
Based on formation energy calculations, among the various clusters with $\bigl[3\mathrm{Ln}_\mathrm{Ca}{:}4\mathrm{F}_i\bigr]^\prime$ composition, the hexameric cluster with  three $\mathrm{Ln}_\mathrm{Ca}^\bullet$ arranged in a semi-circular pattern (Fig.~\subref*{cluster34}) is the most stable structure. For the case of Yb$^{3+}$/Ln$^{3+}$ co-doping, the structure in which $\mathrm{Ln}_\mathrm{Ca}^\bullet$ is located between two $\mathrm{Yb}_\mathrm{Ca}^\bullet$ is the most stable, and is denoted as $\bigl[1\mathrm{Ln}_\mathrm{Ca}{:}2\mathrm{Yb}_\mathrm{Ca}{:}4\mathrm{F}_i\bigr]^\prime$ in Fig.~\subref*{concentration}. For $\bigl[3\mathrm{Ln}_\mathrm{Ca}{:}3\mathrm{F}_i\bigr]^\times$, the structure formed by $C_{4v}$ aggregation (Fig.~\subref*{cluster33}) is the most stable. And the structure for Yb$^{3+}$/Ln$^{3+}$ co-doping is denoted as $\bigl[1\mathrm{Ln}_\mathrm{Ca}{:}2\mathrm{Yb}_\mathrm{Ca}{:}3\mathrm{F}_i\bigr]^\times$.

  While our synthesis was conducted under controlled conditions to minimize oxygen exposure, we acknowledge that trace O$^{2-}$ incorporation is a common concern in fluorides. Charge compensation via an O$^{2-}$ ion at a F$^-$ site (forming $\mathrm{O_F^\prime}$) is a plausible alternative mechanism, as evidenced in our previous studies on related systems (e.g., Ce$^{3+}$-doped CaF$_2$, see Ref.~\cite{jp305593h}). In this study, the consistent spectroscopic and energetic trends across the lanthanide series suggest that Ln$^{3+}$–F$^-$ interactions and Ln$^{3+}$ aggregation are the dominant clustering mechanisms. Intentional oxygen introduction could be a focus of future work to systematically elucidate its role.

\subsection{Concentration of clusters}

Figure~\subref*{ratio} shows that the clustering behavior is affected by the activator, \textit{i.e.}, 2~mol\% Ln$^{3+}$ ion across the lanthanide series. In nanoparticles, the ratio of cluster 1 to $\mathrm{Yb}_\mathrm{Ca}^\bullet$ ($O_h$) cubic decreases with the lanthanide series, whereas the ratio of cluster 2 to cubic is relatively constant. 

To simulate the results, we calculated cluster concentrations for various lanthanide co-doping configurations through DFT-calculated formation energies. The equilibrium temperature needed in the calculation was set to 463.5~K (190~$^\circ$C) to simulate the synthesis process~\cite{BALABHADRA2020155165}. And the relative concentrations are largely unaffected by the equilibrium temperature adopted in the calculation. 

Previous experiments have shown that clustering of Yb$^{3+}$ ions is markedly less pronounced in nanocrystals compared to bulk crystals~\cite{BALABHADRA2020155165}. This is likely due to lower synthesis temperatures, surface effects, and distinct nanoscale phenomena that suppress ion mobility and aggregation. Accordingly, our simulations excluded clusters containing four or more lanthanide ions. Since the DFT calculations based on the supercell method inherently model doping defects in bulk crystalline environments, incorporating such configurations would energetically favor the formation of large hexameric clusters. This is consistent with experimental observations in bulk materials but contradicts the experimental evidence in nanoparticles.

The dominant cluster concentrations for clusters containing up to three lanthanide ions are presented in Fig.~\subref*{concentration}.
The concentration of dominant clusters shows no significant variations along the range of light lanthanide co-dopants. However, for heavier Ln$^{3+}$ ions, the concentration of neutral $\bigl[1\mathrm{Ln}_\mathrm{Ca}{:}2\mathrm{Yb}_\mathrm{Ca}{:}3\mathrm{F}_i\bigr]^\times$ clusters progressively decreases across the lanthanide series. Neutral $\bigl[3\mathrm{Yb}_\mathrm{Ca}{:}3\mathrm{F}_i\bigr]^\times$ clusters are not shown as their concentration is calculated to be negligible. 
The positively charged $\mathrm{Yb}_\mathrm{Ca}^\bullet$ (cubic $O_h$) and negatively charged $\bigl[3\mathrm{Yb}_\mathrm{Ca}{:}4\mathrm{F}_i\bigr]^\prime$ are dominant, and their proportions are relatively constant across the entire lanthanide series. These two species represent the principal positively and negatively charged defects, respectively, and effectively compensate each other in charge. 
When normalized to the cubic $O_h$ center ($\mathrm{Yb}_\mathrm{Ca}^\bullet$), the concentration ratio of negatively charged $\bigl[3\mathrm{Yb}_\mathrm{Ca}{:}4\mathrm{F}_i\bigr]^\prime$ remains constant, while the ratio of the neutral $\bigl[1\mathrm{Ln}_\mathrm{Ca}{:}2\mathrm{Yb}_\mathrm{Ca}{:}3\mathrm{F}_i\bigr]^\times$ cluster decreases across the $\mathrm{Ln}$ series. This trend aligns with the decreasing tendency to form a $C_{4v}$ monomer, which arises from reduced structure compatibility between the $\mathrm{F_8}$ cage containing $\mathrm{F}_i^{\prime}$ and its adjacent cage containing $\mathrm{Ln_{Ca}^{\bullet}}$, whose size decreases across the lanthanide series due to the progressive reduction in ionic radius~\cite{CAI2022119058}.
  Table S3 in Supplementary Material presents the relaxed Ln–F bond lengths (serving as a proxy for ionic radii in eight-fold coordination within our model) for the $\mathrm{Ln_{Ca}^{\bullet}}$ defect with $O_h$ symmetry across the lanthanide series, providing a consistent internal reference for interpreting the observed trends.

Comparison between the DFT calculations and experimental measurements are complicated by the fact that the absorption coefficients of different sites will vary. However, comparing the variations across the series of the calculated concentrations and experimental measurements, 
{we tentatively conclude that cluster peak 1 at $\sim$10225~cm$^{-1}$ is dominated by neutral $C_{4v}$ aggregates containing Ln$^{3+}$, such as $\bigl[1\mathrm{Ln}_\mathrm{Ca}{:}2\mathrm{Yb}_\mathrm{Ca}{:}3\mathrm{F}_i\bigr]^\times$, while cluster peak 2 at $\sim$10265~cm$^{-1}$ is dominated by negatively charged derivatives of hexameric cluster, such as $\bigl[3\mathrm{Yb}_\mathrm{Ca}{:}4\mathrm{F}_i\bigr]^\prime$.

In our previous work on infrared to visible upconversion in Yb$^{3+}$/Er$^{3+}$ co-doped CaF$_2$ nanocrystals~\cite{BALABHADRA2020155165}, we showed that the upconversion was not proportional to the absorption coefficient. From Fig.~6(a) of Ref.~\cite{BALABHADRA2020155165} we see that the cluster 1 region has a higher upconversion signal relative to the absorption cross-section than the cluster 2 region. This is consistent with the cluster 1 region being dominated by neutral $\bigl[1\mathrm{Ln}_\mathrm{Ca}{:}2\mathrm{Yb}_\mathrm{Ca}{:}3\mathrm{F}_i\bigr]^\times$ clusters, for which upconversion will be efficient and the cluster 2 region being dominated by negative $\bigl[3\mathrm{Yb}_\mathrm{Ca}{:}4\mathrm{F}_i\bigr]^\prime$ clusters without Ln$^{3+}$, for which upconversion will be less efficient.

In this work we have used DFT calculations and FTIR spectroscopy to gain insight into heterogeneous clustering in the nanophase. A full characterization of cluster formation would involve significantly more computational complexity. A complete experimental identification of the various cluster configurations would be even more challenging. It would require systematic synthesis of diverse samples, characterization of the Ln$^{3+}$ and Yb$^{3+}$ cations by a variety of spectroscopic methods, and detailed modelling to establish the structures of the clusters.

\section{Conclusions}

In conclusion, we have investigated the formation and characteristics of heterogeneous lanthanide ion clusters in CaF$_2$ nanocrystalline materials, by combining experimental and computational approaches. CaF$_2$ nanoparticles co-doped with 20~mol\% Yb$^{3+}$ and 2~mol\% Ln$^{3+}$ (Ln$^{3+}$ = Ce$^{3+}$, Pr$^{3+}$, Nd$^{3+}$, Sm$^{3+}$, Eu$^{3+}$, Gd$^{3+}$, Ho$^{3+}$, Er$^{3+}$, and Tm$^{3+}$) were synthesized via a hydrothermal method. Their structural and morphological properties of the nanoparticles were characterized using PXRD, DLS, and TEM techniques, confirming their cubic phase and nanoscale size.

High-resolution FTIR spectroscopy was employed to study the absorption spectra of the nanoparticles, revealing the presence of Yb$^{3+}$ single ion cubic centers and various cluster sites. 
The experiments demonstrated that the concentrations of different clusters is affected by the the co-doped Ln$^{3+}$ ion, even though the concentration of Yb$^{3+}$ is much larger than that of the co-dopant.

DFT calculations were performed to estimate the formation energies and local coordination structures of different clusters. The calculated trends of cluster concentrations are consistent with the experimental observations and provide insight into the clustering behavior in the nanoparticles. The results indicate that the neutral $C_{4v}$ aggregations containing Ln$^{3+}$, such as $\bigl[1\mathrm{Ln}_\mathrm{Ca}{:}2\mathrm{Yb}_\mathrm{Ca}{:}3\mathrm{F}_i\bigr]^\times$ clusters, tend to decrease across the lanthanide series, as the Ln$^{3+}$ ions reduce in size. In contrast, the concentrations of negatively charged derivatives of the hexameric cluster, such as $\bigl[3\mathrm{Yb}_\mathrm{Ca}{:}4\mathrm{F}_i\bigr]^\prime$ clusters, remain relatively constant across the lanthanide series. 

This study advances the understanding of clustering mechanisms in lanthanide-doped CaF$_2$ nanoparticles and their impact on luminescence properties such as upconversion. The results may guide the design of advanced nanomaterials with improved luminescence performance for applications in bioimaging, thermometry and lighting.

\section*{Declaration of competing interest}
The authors declare that they have no known competing financial interests or personal relationships that could have appeared to influence the work reported in this paper.

\section*{Acknowledgment}
S.B. and H.X. are co-first authors and contributed equally to this work.

The authors gratefully acknowledge the technical support provided by Mr. Stephen Hemmingson, Mr. Graeme MacDonald, Dr. Matthew Polson, and Mr. Robert Thirkettle of the University of Canterbury, New Zealand. 

This work was supported by the National Natural Science Foundation of China under grant no.\ 12304445 (J.C.) and no.\ 12474242 (H.X. and C.K.D.). The numerical calculations were performed on the supercomputing system in the Supercomputing Center of University of Science and Technology of China. Coordination structures were visualized using VESTA~\cite{Momma:db5098}.

\bibliographystyle{elsarticle-num} 
\bibliography{ref}

\clearpage


\section*{Supplementary material}

# Supplementary Material

## Lanthanide-Dependent Clustering in Yb$^{3+}$/Ln$^{3+}$ Co-Doped CaF$_2$ Nanocrystals: Correlating Spectroscopic Signatures with DFT Insights

Sangeetha Balabhadra, Haoming Xu, Jiajia Cai, Chang-Kui Duan, Michael F. Reid, Jon-Paul R. Wells

**Table S1:** Formation energy (in eV) of various clusters (as well as point defects) for Yb singly doped CaF2 crystals.

| Composition | Cluster structure description | Formation energy |
|---|---|---|
| $V_{Ca}''$ | Ca vacancy | 1.658 |
| $V_F^{\cdot}$ | F vacancy | 1.114 |
| $F_i'$ | F interstitial | 0.989 |
| $Yb_{Ca}^{\cdot}$ | One Yb replaces Ca site | 0.113 |
| $[1Yb_{Ca} \cdot 1F_i]^{\times}$ | One Yb replaces Ca site, accompanied by a nearest F interstitial (C4v) | 0.778 |
| | One Yb replaces Ca site, accompanied by a second nearest F interstitial (C3v) | 0.751 |
| $[2Yb_{Ca} \cdot 1F_i]^{\cdot}$ | Two Yb are adjacent to each other | 3.762 |
| | Two Yb are second nearest neighbors | 6.412 |
| $[2Yb_{Ca} \cdot 2F_i]^{\times}$ | Two Yb adjacent to each other | 0.669 |
| | Two Yb are second nearest neighbors | 1.312 |
| $[2Yb_{Ca} \cdot 3F_i]'$ | Two Yb are adjacent to each other | 8.948 |
| | Two Yb are second nearest neighbors | 10.234 |
| $[3Yb_{Ca} \cdot 2F_i]^{\cdot}$ | Three Yb are adjacent to each other | 6.938 |
| | One Yb is adjacent to two second nearest neighbors Yb separately | 7.969 |
| $[3Yb_{Ca} \cdot 3F_i]^{\times}$ | Three Yb are adjacent to each other | 0.305 |
| | One Yb is adjacent to two second nearest neighbors Yb separately | 1.881 |
| | Three Yb are adjacent to each other, forming a hexamer together with three Ca | 1.378 |
| | One Yb is adjacent to two second nearest neighbors Yb separately, forming a hexamer together with three Ca | 1.184 |
| $[3Yb_{Ca} \cdot 4F_i]'$ | Three Yb are adjacent to each other | 0.258 |
| | One Yb is adjacent to two second nearest neighbors Yb separately | 2.296 |
| | Three Yb are adjacent to each other, forming a hexamer together with three Ca | 0.271 |
| | One Yb is adjacent to two second nearest neighbors Yb separately, forming a hexamer together with three Ca | 0.212 |

Notes: The chemical potentials for $F^-$, $Ca^{2+}$, and $Yb^{3+}$ are -2.574 eV, -12.803 eV, and -19.612 eV, respectively, determined via Eqs. (3), (5), and (6) in the main text. The structural screening was conducted following the protocol from Reference [43], which is based on our prior findings.

**Table S2:** The chemical potentials and formation energies (in eV) of the clusters for different co-doping lanthanides.

| Co-doping Ln | Ce | Pr | Nd | Pm | Sm | Eu | Gd | Tb | Dy | Ho | Er | Tm |
|---|---|---|---|---|---|---|---|---|---|---|---|---|
| $\mu(F^-)$ | -2.570 | -2.570 | -2.570 | -2.570 | -2.570 | -2.570 | -2.569 | -2.569 | -2.569 | -2.568 | -2.569 | -2.572 |
| $\mu(Ca^{2+})$ | -12.811 | -12.811 | -12.811 | -12.811 | -12.811 | -12.811 | -12.812 | -12.812 | -12.813 | -12.814 | -12.812 | -12.806 |
| $\mu(Yb^{3+})$ | -19.627 | -19.627 | -19.627 | -19.627 | -19.627 | -19.627 | -19.628 | -19.629 | -19.630 | -19.631 | -19.628 | -19.617 |
| $\mu(Ln^{3+})$ | -19.700 | -19.806 | -19.858 | -19.901 | -19.862 | -19.875 | -19.875 | -19.839 | -19.793 | -19.748 | -19.709 | -19.713 |
| $V_{Ca}''$ | 1.650 | 1.650 | 1.650 | 1.650 | 1.650 | 1.650 | 1.650 | 1.649 | 1.648 | 1.647 | 1.649 | 1.655 |
| $V_F^{\cdot}$ | 0.993 | 0.993 | 0.993 | 0.993 | 0.993 | 0.993 | 1.110 | 1.110 | 1.109 | 1.109 | 1.109 | 1.113 |
| $F_i'$ | 1.110 | 1.110 | 1.110 | 1.110 | 1.110 | 1.110 | 0.993 | 0.993 | 0.994 | 0.994 | 0.993 | 0.990 |
| $Yb_{Ca}^{\cdot}$ | 0.120 | 0.120 | 0.120 | 0.120 | 0.120 | 0.120 | 0.121 | 0.121 | 0.121 | 0.121 | 0.120 | 0.116 |
| $Ln_{Ca}^{\cdot}$ | 0.474 | 0.447 | 0.416 | 0.385 | 0.356 | 0.319 | 0.280 | 0.249 | 0.219 | 0.195 | 0.184 | 0.185 |
| $C_{4v}$ [1Yb$_{Ca}$·1F$_i$]$^{\times}$ | 0.782 | 0.782 | 0.782 | 0.782 | 0.782 | 0.782 | 0.782 | 0.782 | 0.781 | 0.781 | 0.780 | 0.780 |
| $C_{3v}$ [1Yb$_{Ca}$·1F$_i$]$^{\times}$ | 0.755 | 0.755 | 0.755 | 0.755 | 0.755 | 0.755 | 0.755 | 0.755 | 0.755 | 0.754 | 0.754 | 0.753 |
| $C_{4v}$ [1Ln$_{Ca}$·1F$_i$]$^{\times}$ | 0.776 | 0.784 | 0.787 | 0.790 | 0.790 | 0.787 | 0.782 | 0.779 | 0.776 | 0.779 | 0.792 | 0.825 |
| $C_{3v}$ [1Ln$_{Ca}$·1F$_i$]$^{\times}$ | 1.230 | 1.191 | 1.147 | 1.099 | 1.062 | 1.014 | 0.965 | 0.923 | 0.884 | 0.855 | 0.835 | 0.829 |
| [2Yb$_{Ca}$·1F$_i$]$^{\cdot}$ | 3.774 | 3.774 | 3.774 | 3.774 | 3.774 | 3.774 | 3.774 | 3.774 | 3.774 | 3.773 | 3.772 | 3.767 |
| [1Ln$_{Ca}$·1Yb$_{Ca}$·1F$_i$]$^{\cdot}$ | 4.404 | 3.963 | 4.035 | 4.112 | 4.015 | 4.030 | 4.271 | 4.126 | 4.298 | 4.019 | 3.837 | 3.848 |
| [2Yb$_{Ca}$·2F$_i$]$^{\times}$ | 0.677 | 0.677 | 0.677 | 0.677 | 0.677 | 0.677 | 0.677 | 0.677 | 0.677 | 0.676 | 0.675 | 0.673 |
| [1Ln$_{Ca}$·1Yb$_{Ca}$·2F$_i$]$^{\times}$ | 0.863 | 0.853 | 0.835 | 0.818 | 0.804 | 0.782 | 0.783 | 0.761 | 0.742 | 0.733 | 0.725 | 0.737 |
| [2Yb$_{Ca}$·3F$_i$]$'$ | 8.952 | 8.952 | 8.952 | 8.951 | 8.951 | 8.951 | 8.951 | 8.951 | 8.950 | 8.948 | 8.948 | 8.950 |
| [1Ln$_{Ca}$·1Yb$_{Ca}$·3F$_i$]$'$ | 0.540 | 0.539 | 0.538 | 0.538 | 0.540 | 0.534 | 0.532 | 0.532 | 0.529 | 0.530 | 0.541 | 0.559 |
| [3Yb$_{Ca}$·3F$_i$]$^{\times}$ | 0.316 | 0.316 | 0.316 | 0.316 | 0.316 | 0.316 | 0.316 | 0.316 | 0.315 | 0.314 | 0.312 | 0.311 |
| [1Ln$_{Ca}$·2Yb$_{Ca}$·3F$_i$]$^{\times}$ | 0.273 | 0.273 | 0.273 | 0.273 | 0.274 | 0.274 | 0.275 | 0.277 | 0.281 | 0.291 | 0.307 | 0.345 |
| [3Yb$_{Ca}$·4F$_i$]$'$ | 0.220 | 0.220 | 0.220 | 0.220 | 0.220 | 0.219 | 0.219 | 0.218 | 0.217 | 0.216 | 0.215 | 0.216 |
| [1Ln$_{Ca}$·2Yb$_{Ca}$·4F$_i$]$'$ | 1.172 | 1.072 | 0.977 | 0.873 | 0.800 | 0.705 | 0.618 | 0.540 | 0.471 | 0.405 | 0.358 | 0.321 |

Notes: Considering that the amount of Ln is relatively small compared to Yb when 20% Yb and 2% Ln are co-doped, only one Ln atom is added into the stable cluster during the construction of the co-doped clusters. The chemical potentials for $F^-$, $Ca^{2+}$, $Yb^{3+}$, and $Ln^{3+}$ are determined via Eqs. (3), (5), (6) and (7) in the main text.

**Table S3:** Ionic radii (in Å) of the lanthanide ions in eight-fold coordination.

| Ln | Ce | Pr | Nd | Pm | Sm | Eu | Gd |
|---|---|---|---|---|---|---|---|
| Ln-F | 2.397 | 2.379 | 2.365 | 2.348 | 2.336 | 2.324 | 2.311 |
| Ln | Tb | Dy | Ho | Er | Tm | Yb | Lu |
| Ln-F | 2.302 | 2.292 | 2.283 | 2.276 | 2.264 | 2.255 | 2.248 |

Notes: The table presents the relaxed coordination bond lengths of the $Ln_{Ca}^{\cdot}$ defect with $O_h$ symmetry. For reference, the Ca–F bond length in perfect $CaF_2$ is 2.385 Å, while the distance from the relaxed interstitial $F^-$ ion to other $F^-$ ions in the F interstitial defect is 2.538 Å.



\end{document}